\documentclass[]{emulateapj}

\newcommand {\omegape}	{\omega_\mathrm{pe}}

\shorttitle{COLLISIONLESS SHOCKS IN UNMAGNETIZED PLASMAS}
\shortauthors{KATO AND TAKABE}

\begin{document}

\title{Non-relativistic Collisionless Shocks in Unmagnetized Electron-Ion Plasmas}

\author{Tsunehiko N. Kato and Hideaki Takabe}
\affil{
Institute of Laser Engineering, Osaka University,
2-6 Yamada-oka, Suita, Osaka 565-0871, Japan
}
\email{kato-t@ile.osaka-u.ac.jp}

\begin{abstract}
We show that the Weibel-mediated collisionless shocks
are driven at non-relativistic propagation speed ($0.1c < V < 0.45c$)
in unmagnetized electron-ion plasmas
by performing two-dimensional particle-in-cell simulations.
It is shown that
the profiles of the number density and the mean velocity
in the vicinity of the shock transition region,
which are normalized by the respective upstream values,
are almost independent of the upstream bulk velocity,
i.e., the shock velocity.
In particular,
the width of the shock transition region
is $\sim 100$ ion inertial length
independent of the shock velocity.
For these shocks
the energy density of the magnetic field generated by the Weibel-type instability
within the shock transition region
reaches typically 1-2\% of the upstream bulk kinetic energy density.
This mechanism probably explains the robust formation of collisionless shocks,
for example, driven by young supernova remnants,
with no assumption of external magnetic field in the universe.
\end{abstract}


\section{Introduction}
Collisionless shocks are important phenomena in the universe;
they heat and compress interstellar or intergalactic plasmas
and often accelerate charged particles or cosmic rays.
The dissipation mechanism of the collisionless shocks is
highly complex and generally involved with
instabilities in the collisionless plasma.

The Weibel instability is driven in anisotropic or
counter-streaming collisionless plasmas \citep{Weibel59, Fried59}
and generates strong magnetic fields.
Collisionless shocks
inevitably generate an anisotropy
within their transition region,
where the upstream and downstream plasmas are mixed up,
therefore, the Weibel instability can develop
and generate strong magnetic fields there
if the background magnetic field is weak.
\citet{Moiseev63} considered such a process in non-relativistic
collisionless shocks and
\citet{Medvedev99} discussed the magnetic field generation
by the Weibel instability in relativistic collisionless shocks
of gamma-ray bursts.

In this context,
the Weibel instability in relativistic counter-streaming plasmas
has been investigated extensively by means of particle-in-cell simulations
\citep[e.g.,][]{Kazimura98, Gruzinov01, Haruki03, Silva03, Nishikawa03, Frederiksen04,
Spitkovsky05}.
Recent research with simulations showed that
the ``Weibel-mediated'' collisionless shocks can form
at relativistic propagation speed
in electron-positron plasmas \citep{Kato07,Chang08}
and also in electron-ion plasmas \citep{Spitkovsky08}
without background magnetic field;
in these shocks the dissipation mechanism is
effectively provided by the magnetic fields
generated by the Weibel-type instability
in the shock transition region.
In the universe,
these shocks can be driven 
associated with relativistic phenomena, such as
gamma-ray bursts, jets from active galactic nuclei,
or pulsar winds.

There is also a possibility that
these Weibel-mediated shocks are driven
even in non-relativistic phenomena,
e.g.,
shocks of supernova remnants
whose propagation speed is typically $\sim 1,000-10,000$ km s$^{-1}$.
However,
in this case,
the Weibel instability can be inefficient
because the linear growth rate of the instability
is proportional to the flow velocity $V$.
Therefore,
it is not clear yet whether this kind of shocks can exist
in non-relativistic regime.
In this paper,
however,
we show that the Weibel-mediated collisionless shocks
propagating at non-relativistic speed
can form in unmagnetized electron-ion
plasmas by performing numerical simulations.
This also suggests a possibility to proof the formation of
these shocks in laboratory plasmas 
with large-scale laser facilities.

\section{Method}
In order to investigate the collisionless shocks
in electron-ion plasmas without background magnetic fields,
we performed numerical simulations.
The simulation code is a relativistic and electromagnetic particle-in-cell code
with two spatial and three velocity dimensions (2D3V)
which was developed based on a standard method described by \cite{Birdsall}.
The basic equations are the Maxwell's equations
and the (relativistic) equation of motion of particles.
The simulation plane is the $x-y$ plane.
The $z$ axis is perpendicular to the plane.
Since we consider collisionless shocks in
unmagnetized plasmas,
the electromagnetic fields are initially set to zero
over the entire simulation box.
The boundary condition of the electromagnetic field
is periodic to each direction.

In the simulations
a collisionless shock is driven
by means of the ``injection method.''
There are two rigid walls at the left-hand side (smaller $x$) and right-hand side (larger $x$)
of the simulation box
and these walls reflect particles specularly.
Initially, both electrons and ions are uniformly loaded
in the region between the two walls
at the bulk flow velocity of $V$ in the $+x$-direction.
At the early stage of the simulations,
the particles located near the right wall
were reflected by the wall
and then interacted with the incoming particles,
i.e., the upstream plasma,
via some instabilities.
Such instabilities induce strong electric or magnetic fields that
provide dissipation mechanism of the shock and,
eventually, a collisionless shock is formed.
Note that the frame of the simulation
corresponds to the downstream rest frame.

\section{Results}
We have carried out a series of simulations
with upstream bulk velocities of $V = 0.45c$, $0.2c$, and $0.1c$,
where $c$ is the speed of light.
The thermal velocity of electrons is typically taken as
one-tenth of the bulk velocity
and that of ions is determined so that the temperatures of electrons and ions are same.
Because of the computational power,
we used a reduced ion mass of $m_i = 20 m_e$.
Here,
we take $\omegape^{-1}$ as the unit of time
and the electron skin depth $\lambda_e \equiv c \omegape^{-1}$ 
as the unit of length,
where $\omegape \equiv (4\pi n_{e0} e^2 / m_e)^{1/2}$
is the electron plasma frequency defined for the electron
number density of the upstream plasma, $n_{e0}$.
We also define the ion inertial length, $\lambda_i \equiv (m_i/m_e)^{1/2} \lambda_e$.
The units of electric and magnetic fields are both taken as
$E_* = B_* \equiv c (4\pi n_{e0} m_e)^{1/2}$.

\subsection{Results for $V=0.45c$}
\begin{figure}[h]
\plotone{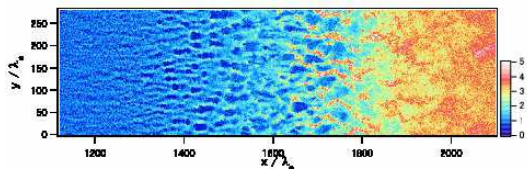}
\caption{
Number density of ions for the upstream bulk velocity of $V = 0.45c$
at $\omegape t = 2100$.
The left-hand side and right-hand side are the upstream and downstream of
the shock, respectively.
Filamentary structure is visible in the shock transition region,
$1300 < x/\lambda_e < 1900$.
}
\label{fig:npV045}
\end{figure}

The simulation for $V = 0.45c$ was performed
on a grid of $N_x \times N_y = 4096 \times 512$
with $\sim 27$ particles per cell per species.
The size of the simulation box is $L_x \times L_y = 2240 \lambda_e \times 280 \lambda_e$.
(We choose the grid size so that the system is stable against the cold beam nonphysical
instability, c.f. \citet{Birdsall}.)
The number density of ions
obtained from the simulation at $\omegape t = 2100$
is shown in Fig.~\ref{fig:npV045}.
The color shows the number density normalized by the upstream value,
namely, $n_{e0}$.
The upstream plasma flows from the left to the right
and goes through the transition region which has a filamentary structure
and then reaches the almost uniform downstream state ($x / \lambda_e > 1900$).

\begin{figure}[h]
\plotone{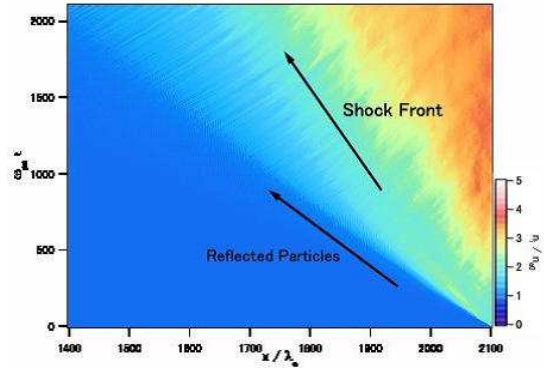}
\caption{
Time evolution of the number density of ions for $V = 0.45c$.
The horizontal and vertical axes are the $x$-coordinate and the time,
respectively.
The color shows the number density averaged over the $y$-direction.
The shock transition region, or ``shock front'', is visible as a steep increase in the number density.
}
\label{fig:npV045dev}
\end{figure}

Figure \ref{fig:npV045dev} shows the time evolution of the number density of ions.
The number density shown in color is averaged over the $y$-direction.
The transition region, or ``shock front'',
which is visible as a steep increase in the number density,
propagates upstream at an almost constant speed
($\sim -0.18c$ measured in the downstream frame)
after $\omegape t \sim 500$.
Note that
the particles that are reflected at the right wall
at the early stage of the simulation,
which are remains of the initial condition,
eventually fade away at later times.

\begin{figure}[h]
\plotone{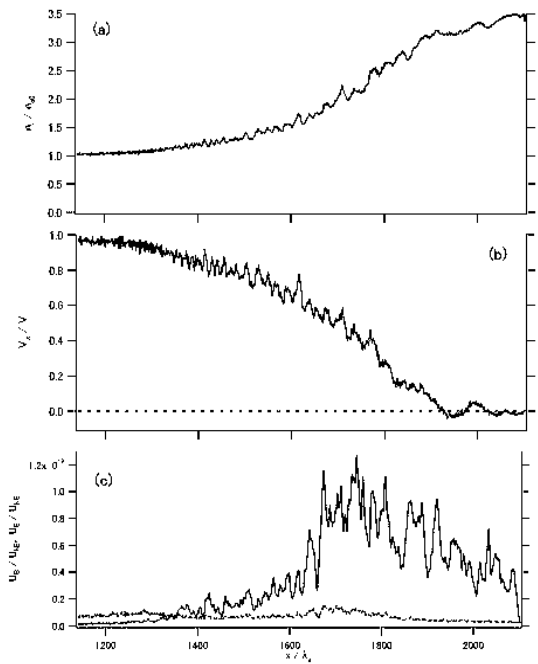}
\caption{
Profiles of (a) the ion number density,
(b) the mean velocity in the $x$-direction
both normalized by the respective upstream values,
and (c) the energy densities of
the magnetic (solid curve) and electric (dashed curve) fields
both normalized by the particle bulk kinetic energy density
of the upstream plasma.
}
\label{fig:profiles_V045}
\end{figure}

Figure \ref{fig:profiles_V045} shows profiles of quantities
averaged over the $y$-direction at $\omegape t = 2100$:
(a) the ion number density, (b) the mean velocity in the $x$-direction,
and (c) the energy densities of electric and magnetic fields.
Both number density and mean velocity rapidly
change to reach the downstream values through the transition region
which extends with the width of $W \sim 500 \lambda_e$.
It is remarkable that a strong magnetic field is generated
within the shock transition region.
The energy density of the magnetic field
reaches approximately $1\%$ of the upstream bulk kinetic energy density
(measured in the downstream rest frame),
$U_\mathrm{KE} = n_{e0} (m_i + m_e) V^2 / 2$.
This strong magnetic field deflects and isotropize the particles coming from upstream;
this provides an effective dissipation mechanism
for this collisionless shock.

\begin{figure}[h]
\plotone{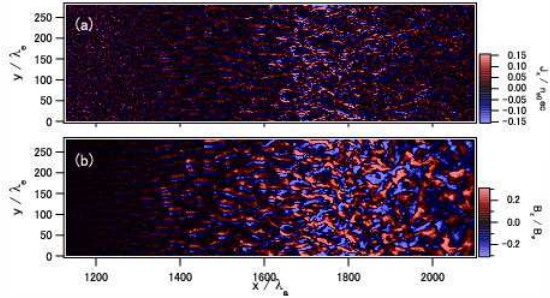}
\caption{
(a) The current density $J_x$
and
(b) the magnetic field $B_z$
in the vicinity of the shock transition region
at $\omegape t = 2100$.
}
\label{fig:JxBz_V045}
\end{figure}

Figure \ref{fig:JxBz_V045} shows the current density
in the $x$-direction, $J_x$, and the $z$-component of the magnetic
field, $B_z$, at $\omegape t = 2100$.
It is evident that a number of current filaments exist
within the transition region.
These filaments consist of ions which are surrounded by the diffuse electrons
similar to the case of the relativistic counter-streaming plasmas \citep{Frederiksen04}.
The filamentary structure seen in the number density in Fig.~\ref{fig:npV045}
reflects the existence of them.
As seen in the figure,
these filaments generate the strong magnetic field around themselves
which is also observed in Fig.~\ref{fig:profiles_V045} (c).
This structure would be formed through the Weibel-type instability
like the relativistic shocks in unmagnetized plasmas \citep{Kato07,Spitkovsky08}.

\subsection{Dependence on upstream bulk velocity and ion mass}

\begin{figure}[h]
\plotone{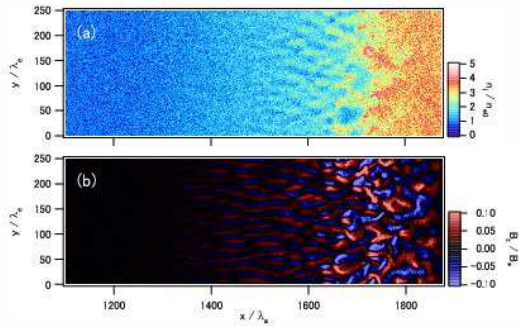}
\caption{
(a) The ion number density and (b) the magnetic field $B_z$
obtained from the simulation for $V = 0.1c$.
}
\label{fig:npBzV01}
\end{figure}

We also performed simulations with the bulk velocities of $0.2c$ and $0.1c$.
In both cases we confirmed that
the structure of the shock is qualitatively identical
to that of $V=0.45c$ case.
The ion number density and the magnetic field
obtained for $V = 0.1c$ are shown in Fig.~\ref{fig:npBzV01}.

\begin{figure}[h]
\plotone{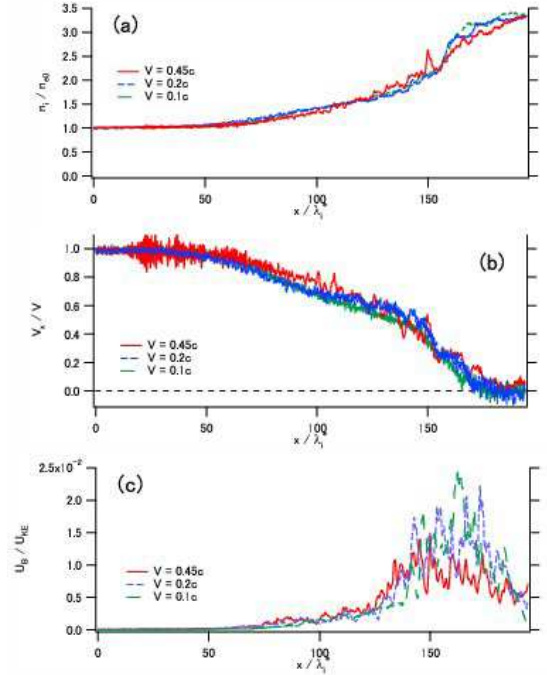}
\caption{
Profiles of
(a) the ion number densities,
(b) the mean velocities in the $x$-direction $V_x$,
and (c) the energy densities of magnetic fields $U_B$
for $V = 0.45c$ (red solid curves), $0.2c$ (blue short-dashed curves),
and $0.1c$ (green long-dashed curves).
The horizontal axes are the $x$-coordinate normalized by the ion inertial length
(including the relativistic effect), $\lambda_i^*$.
The normalizations of the profiles are same as in Fig.~\ref{fig:profiles_V045}.
}
\label{fig:npVprofile}
\end{figure}

It should be noted that these Weibel-mediated shocks
have similarities in the profiles of some quantities
under changing the upstream bulk velocity $V$.
Figure \ref{fig:npVprofile} shows
the profiles as Fig.~\ref{fig:profiles_V045} at different upstream bulk velocities:
$V = 0.45c$, $0.2c$ and $0.1c$.
Here, the horizontal axes are the $x$-coordinate normalized by
the ion inertial length including the relativistic effect,
$\lambda_i^* \equiv \Gamma^{1/2} \lambda_i$,
where $\Gamma \equiv [1 - (V/c)^2]^{-1/2}$ is the Lorentz factor
of the upstream bulk velocity.
It is remarkable that
the profiles in (a) and (b) are almost identical independent of $V$.
In particular,
the widths of the shock transition region are $W \sim 100 \lambda_i^*$
in \textit{all} cases.
The energy densities of magnetic fields
within the transition region
also reach typically 1-2\% of the upstream bulk kinetic energy density at maximum
independent of $V$.

\begin{figure}[h]
\plotone{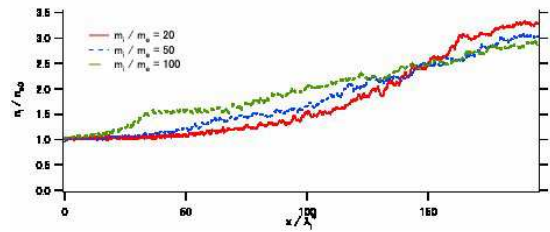}
\caption{
Profiles of the ion number densities
for different ion-to-electron mass ratios:
$m_i/m_e = 20$ (red solid curve), $50$ (blue short-dashed curve),
and $100$ (green long-dashed curve).
The upstream bulk velocities are $V = 0.45c$ in all cases.
The horizontal and vertical axes are same as in Fig.~\ref{fig:npVprofile} (a).
}
\label{fig:M_profile}
\end{figure}

In order to see the effect of the mass ratio,
we performed simulations for $m_i / m_e = 50$ and $100$
with $V = 0.45c$.
The normalized ion number densities for the different three mass ratios
are shown in Fig.~\ref{fig:M_profile} as in Fig.~\ref{fig:npVprofile} (a).
The width of the transition region
is approximately $W \sim 100 \lambda_i^*$ in all cases
and this suggests the width may be the same order even for the real mass ratio.
However, there is a slight difference in the profiles and
this will be investigated in detail in the forthcoming paper.

\section{Discussion}
The similarities in the normalized profiles shown in Fig.~\ref{fig:npVprofile}
can be explained as follows.
The structure of the transition region is essentially
determined by the coalescence history of the ion current filaments.
Here, we simply approximate the coalescence of two filaments as that of
two straight currents.
In this approximation,
the current in each filament is proportional to the bulk velocity $V$,
and a dimensional analysis of the equation of motion of the filaments
shows the time scale of the coalescence of two filaments
due to the Lorentz force, $\tau$,
is proportional to $V^{-1}$.
Thus, if we consider that the width in which the scale of the filamentary structure
becomes twice is given by $\Delta W \sim V \tau$, it is independent of $V$.
On the other hand,
the downstream end of the transition region is determined by
the saturation of the magnetic field of the ion current filaments;
i.e., the typical gyro-radius of ions becomes
comparable with the filament size, $r_{g,i} \sim r_f$,
like the case of electron-positron current filaments \citep{Kato05}.
This gives the condition $r_f \sim \lambda_i$,
which is independent of $V$.
Therefore,
if we consider the initial scale of the ion current filaments
is on the order of the electron skin depth $\lambda_e$,
the total width of the transition region is also independent of $V$.

Though the results described in the previous section
were obtained for unmagnetized plasmas,
these Weibel-mediated shocks can also be driven in magnetized
plasmas if the magnetic field is sufficiently weak.
One of the criteria for neglecting the upstream background magnetic field, $B_0$,
may be given by $r_{g,i} / W \gg 1$,
where $r_{g,i}$ is the typical gyro-radius of ions defined for $B_0$
and $W$ is the typical width of the shock transition region.
If we define the magnetization parameter $\sigma$ as
$\sigma \equiv [B_0^2/8\pi] / [n_{e0} (m_i + m_e) V^2 /2]$
and use the result obtained in the previous section $W \sim 100 \lambda_i^*$,
the criterion can be rewritten $\sigma \ll 10^{-4}$.
For shocks in supernova remnants (SNRs) of $\sim 1000$ years old,
typical parameters are
$n_{e0} \sim 1$ cm$^{-3}$, $V \sim 2000$ km s$^{-1}$, and $B_0 \sim 3\mu$G.
In this case,
the magnetization parameter becomes
$\sigma \sim 10^{-4} - 10^{-5}$
and therefore
there is a possibility that
the Weibel-mediated shocks are driven in SNRs.
For younger SNRs with higher shock speeds of $\ge 10,000$ km s$^{-1}$
\citep[e.g.,][]{Reynolds08},
these shocks would be formed more easily.

If the Weibel-mediated shocks are driven even
at the velocity of $\sim 1000$ km s$^{-1}$,
they can be investigated in laboratories
because laser experiments are capable of generating
collisionless plasma flow at this velocity \citep{Nishimura81, Ripin90}.
For this plasma flow with $n_{e0} \sim 10^{20}$ cm$^{-3}$
(even at this density,
the plasma can be collisionless during the shock formation),
the width of the shock transition region is estimated to be $W \sim 2$ mm.
This is a reasonable value as the plasma size produced with the present-day
large-scale laser facilities.
To demonstrate the shock formation in laboratories
is a challenging attempt
but it would be particularly worthwhile 
as one of the key subjects for laboratory astrophysics
\citep[e.g.,][]{Takabe04}.

\acknowledgments
We would like to thank Y.~Sakawa and Y.~Kuramitsu for their useful discussions
and A.~Kageyama for her helpful comments.
Numerical computations were carried out at Cybermedia Center, Osaka University.


\end{document}